\documentclass{article}

\usepackage{arxiv}

\usepackage[utf8]{inputenc} 
\usepackage[T1]{fontenc}    
\usepackage{hyperref}       
\usepackage{url}            
\usepackage{booktabs}       
\usepackage{amsfonts}       
\usepackage{nicefrac}       
\usepackage{microtype}      
\usepackage{lipsum}
\usepackage{lineno,hyperref}
\usepackage{algorithm}
\usepackage{algorithmic}
\usepackage{tabu}
\usepackage{rotating}
\usepackage{graphicx}
\usepackage[lofdepth,lotdepth]{subfig}
\usepackage{tikz}
 
\usepackage{ragged2e}
\usepackage{multirow}
\usepackage{doi}
\usepackage{amsmath}
\hypersetup{allcolors = red}
\usepackage{array}
\usepackage{tablefootnote}
\usepackage{amsmath}

\title{Generating Trading Signals by ML algorithms or Time Series ones?}

\author{
  Omid Safarzadeh \thanks{Corresponding author} \\
  Department of Economics\\
  Bilkent Unversity\\
  Ankara, Turkey \\
   \texttt{omidpoly@gmail.com} \\
}

\begin{document}
\maketitle

\begin{abstract}
This research investigates efficiency on-line learning Algorithms to generate trading signals. I employed technical indicators based on high frequency stock prices and generated trading signals through ensemble of Random Forests. Similarly, Kalman Filter was used for signaling trading positions. Comparing Time Series methods with Machine Learning methods, results spurious of Kalman filter to Random Forests in case of on-line learning predictions of stock prices.  \end{abstract}


\section{Introduction}
\section{Introduction}
Prediction of stock market prices has a wide literature and is a difficult task. Nowadays, Wall-Street companies employ trading algorithms to process High Frequency stock prices. These algorithms should be able to prepare trading signals (long /short /Do nothing) based on financial technical analysis. Some of these analysis, technical indicators, are derived from stock price fluctuations to signal on the direction of market. In absence of shocks to Market Fundamentals, Wall-Street companies generally use technical analysis to track and trace HF stock movements. This research will examine how ML algorithms can be used to generate trading signals. To do so, several steps needs to be done before apply ML algorithms to data.

\subsection{Data}

In this research, I will use Minuets by Minuets data of several stocks from NYSE. Blomberg Financial station will be the source of stock data. Here is a sample data for Abbott Laboratories company. The data is from 1 Dec. 2017 to 15 Feb. 2018 minutes by minutes intervals. There is 20277 observations of Abbott Laboratories stock price. I will also consider other assets from SP100 stock index.

\begin{table}[ht]
\centering
\begin{tabular}{rlrrrrrrr}
  \hline
 & Date & Open & Close & Low & High & Value & Volume & Number\_Ticks \\ 
  \hline
1 & 12/1/2017 16:31 & 170.01 & 170.40 & 170.01 & 170.71 & 8119385.00 & 47651 & 291 \\ 
  2 & 12/1/2017 16:32 & 170.38 & 170.67 & 170.37 & 170.69 & 5024126.00 & 29450 & 207 \\ 
  3 & 12/1/2017 16:33 & 170.68 & 170.39 & 170.36 & 170.95 & 5591862.00 & 32758 & 220 \\ 
  4 & 12/1/2017 16:34 & 170.42 & 170.42 & 170.28 & 170.50 & 4946603.50 & 29027 & 221 \\ 
  5 & 12/1/2017 16:35 & 170.44 & 170.50 & 170.30 & 170.57 & 5864560.50 & 34406 & 184 \\ 
  6 & 12/1/2017 16:36 & 170.50 & 170.47 & 170.24 & 170.50 & 4071876.25 & 23904 & 150 \\ 
   \hline
\end{tabular}
\end{table}
 \begin{figure}[!htb]
 \frame{\includegraphics[width=15cm]{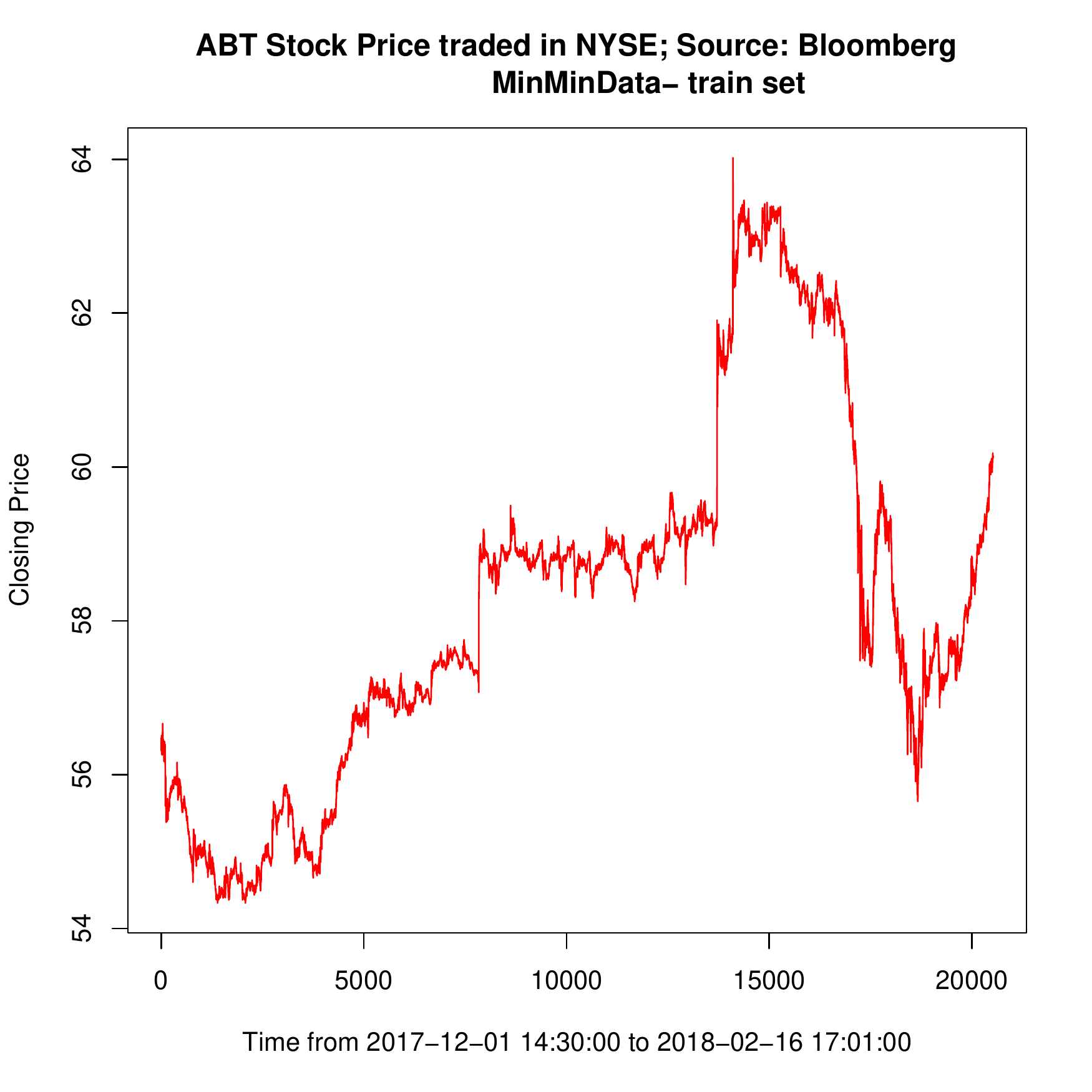}}
\label{Fig:Data16}  
\caption{Abbott Laboratories
NYSE: ABT }
\end{figure}
\hfill

\subsection{Technical Indicators}
The above mentioned data needs to be processed to generate some information about direction of markets. Generally, this will need technical analysis with some parameter calibration to extract proper information of historical behavior of financial asset. It is very common in literature to use financial indicators to provide more information on direction of stock prices,  (see \cite{murphy1999technical} , \cite{teixeira2010method}). These technical indicators are calculated based on daily trading data (open, high, low and close prices) to provide patterns and trends of stock behavior in the incoming trading days. Main stream finance literature ignores them, due to their lack of mathematical foundations following asset pricing models. \cite{godfrey1964random} believed that stock prices are following random walk process, leading that future prices are independent than todays prices; implying no memory price time series. Efficient Market Hypothesis \cite{fama1965behavior} supports random walk theory of prices by stating that, stock prices already include all information about stock value and only new information will change the price. But these theories are violated by some financial economist scholars by providing evidence of market anomalies (\cite{bondt1985does} and \cite{jensen1978some}). Hence, analyzing future prices based on historical data can be seen as another empirical evidence of anomalies in markets. 

\cite{lui1998use} survey analysis revealed that almost 85 \% of foreign exchange traders use both technical and fundamental analysis. They also found that technical analysis is the favorite tool of market traders. The use of technical indicators, ease the complexity of market prediction and transfers it to a pattern recognition problem. we can use technical indicators to provide some features of future market directions. This data transformation, will also eliminates the needs of human interpretation of indicators and helps to implement automated trading algorithm. This transformation is explained in details by \cite{creamer2010automated}.  \\

\subsection{Technical Indicators Data}
Following \cite{kara2011predicting} paper, I will use the following technical indicators as features of daily stock prices and their parameters (see  \cite{creamer2010automated} for more details on parameters adjustments):
\begin{itemize}
\item CCI (Commodity Channel Index)
\item Simple Moving Average (SMA) where $n= 10, 16 , 22$
\item Exponential Moving Average (EMA) where $\lambda = 0.9, 0.84, 0.78$
\item Moving Average Convergence Divergence (MACD)  
\item Momentum: change in prices ($P^t_c$)  for $n$ periods. When it is above (below) zero, indicates is upward
(downward) trend. $n = 12, 18, 24$
\item Relative Strength Index (RSI) where $n1 = 8, 14,  20$
\item Bollinger Bands (BB) where $s=2, n=20, 26 ,32$
\item Chaikin Oscillator 
\item Stochastic Oscillators
\end{itemize}
 \begin{figure}[!htb]
 \frame{\includegraphics[width=18cm]{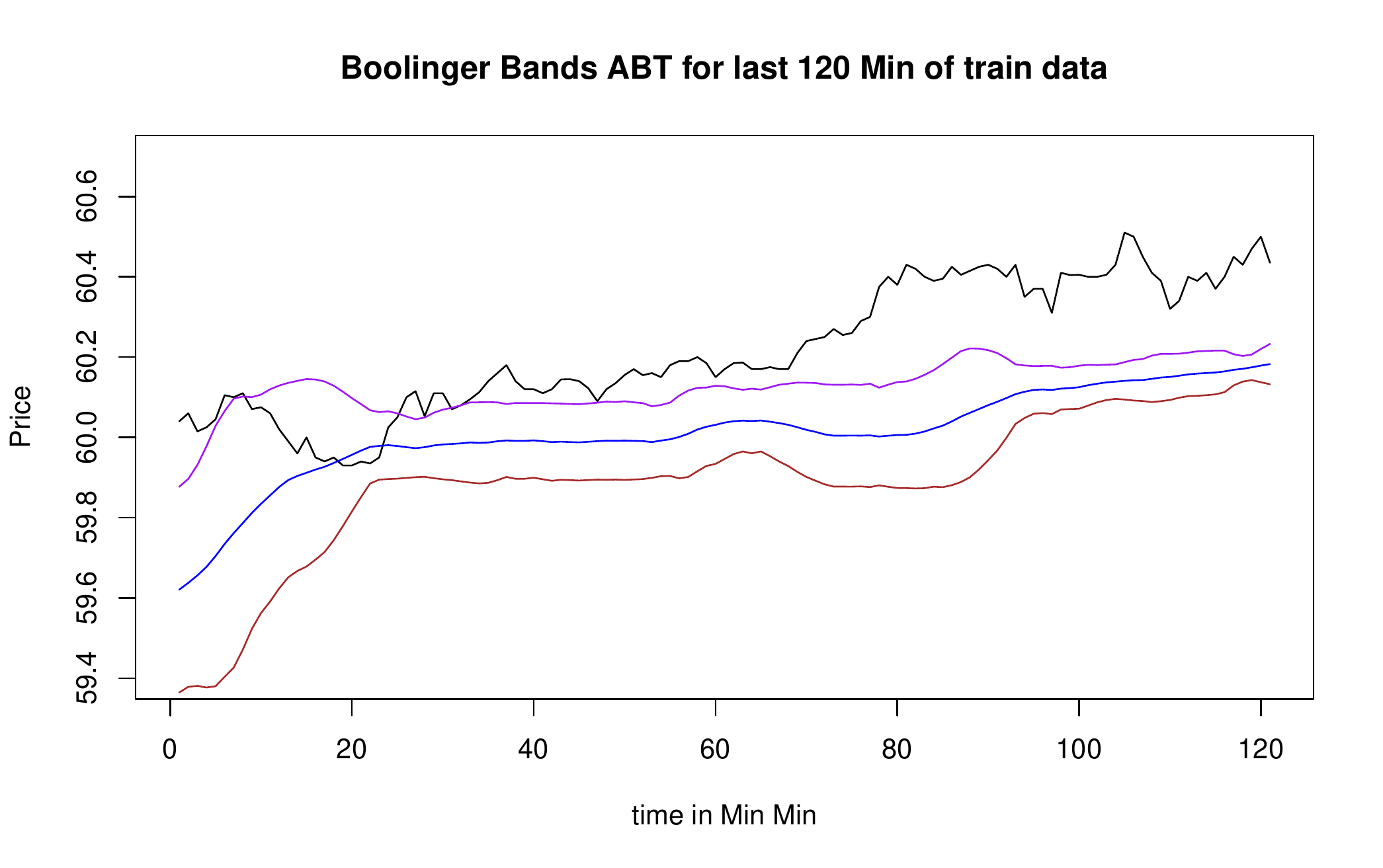}}
\label{Fig:Data11}  
\caption{Bolinger Bands financial technical indicators applied to ABT prices }
\end{figure}
\hfill 

For each of above indicators, there are some parameters such as number of day lags of stock and so on. Here I will use parameters adjustments proposed in \cite{creamer2010automated}, \cite{kara2011predicting} and \cite{booth2015performance}. Using these features, I could provide almost 90 features for each trading days. 

\section{Trading Signals}
Once, the raw data is processed and proper features are extracted to signal the direction of stock prices, the question is what position we should take today, to collect proper profit tomorrow. This means that we have three actions for today state which has some values in tomorrow state. These trading signals will be Long (Buy), Short (Sell) , and "Do Nothing". I am adding the third one as it will take into account some market frictions. Generally, it is common in literature to not consider market frictions such as tax, transaction costs and etc. But introducing the above three actions will be useful when the tomorrow's return is not enough if investors have to also cover market frictions. Hence if prediction of market direction signals a bull market, we long the stock and if it is bear market we short the position. 
Here, I will also consider closing long (or short) positions the day after we opened the position. So our trading strategy will be as:
\begin{eqnarray}
| \hat{X_t}-X_{t-1}|-TAX=0 \Rightarrow No Trade \\
if \quad | \hat{X_t}-X_{t-1}|-TAX>0 \Rightarrow Then: \\
Sign( \hat{X_t}-X_{t-1})=1 \Rightarrow Long(Buy)\\
Sign( \hat{X_t}-X_{t-1})=-1 \Rightarrow Short(Sell)\\
\end{eqnarray}

This means, we will execute the trade if the predicted profit for next minute is higher than market friction. Then depending on market direction we will open long/short position and we will close the position the next minute. 

\section{Stationarity Test}
Time series data are prone to have "random sampling" problem as the order of data is important. In addition, due to external shocks of markets, the underlying random process might not have stationary distribution. So one needs to be careful of using ML algorithms as shuffling data will cost losing auto correlation of data. This means non-stationarity of distribution of data generating process will contradict the assumption of identically distribution of data.
This justifies stationarity test of distribution of data before approaching any prediction procedure. It is well-known that stock prices are random walk and also exhibit auto correlation in time series analysis. Detecting such behavior is possible by using unit root tests in Times Series analysis. This is tested via augmented Dickey Fuller test (ADF)  which is well known in Econometrics literature (see \cite{dickey1979distribution}). Appendix A, provides the R-codes and sample ADF test report for one security. This test is applied for all above mentioned stock prices and similar results have been found. Moreover, density plot of log returns are also demonstrated. The  Kolmogorov–Smirnov test applied to detect the probability distribution of log-return times series, and result was close to skewed-Laplace time series. The green plot is empirical density of series and red lines shows skewed Laplace distribution fitted to data.   
 \begin{figure}[!htb]
 \frame{\includegraphics[width=18cm]{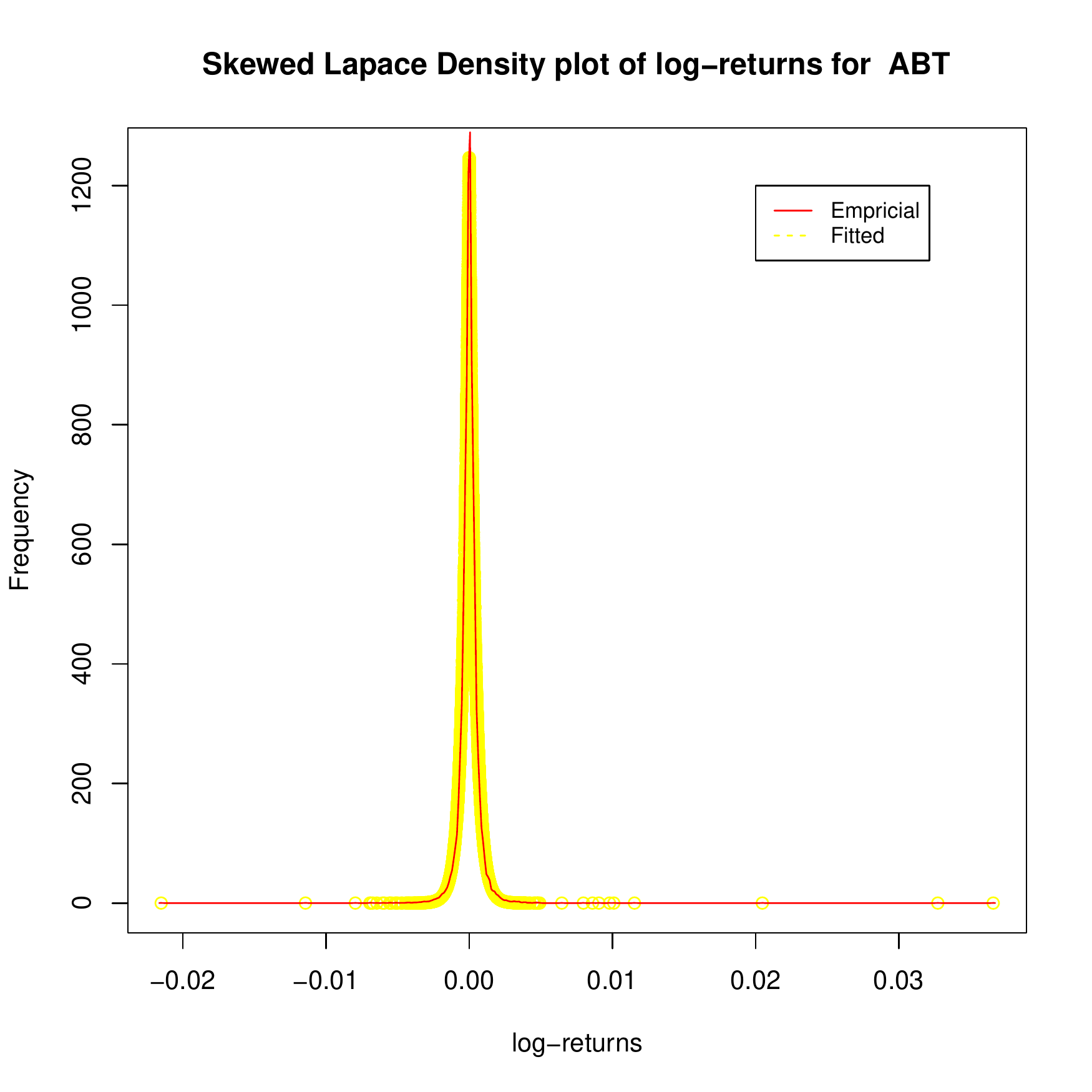}}
\label{Fig:Data12}  
\caption{Skewed Laplace Density plot fitted to log returns of Abbott Laboratories
NYSE: ABT }
\end{figure}
\hfill 
\clearpage
\section{Methodology}

Two approaches has been explored by scholars to predict stock market prices: Time Series analysis and Machine Learning. The former approach considers Financial data as a linearly correlated time dependent information and tries to explain tomorrow prices based on historical data. Machine learning techniques try to capture nonlinear relationship of high dimensional stock prices. Neural Networks, Random Forests and Support Vector Regressions has been widely applied to forecast market prices in literature. \cite{gately1995neural} explains in details how to construct, use and apply neural networks to predict stock prices. \cite{cavalcante2016computational} surveys on recent developments of Machine learning techniques which are used to forecast financial markets. \\

\section{Performance Measure}
The main question of this research is which algorithm we should use to predict and generate trading signals? Which one has superior results? Well, this won't be difficult if we look from investor's viewpoint. Investors don't care about methodology but their profit within a given period. Let's call explained methods in previous section as an Expert. So we are facing two expert types: Time Series and Machine Learning ones. As we are using High Frequency data, we need to define a proper Test time interval. Although it is rule of thumb to split data into 80-20 groups of data as test and train data. Because of sensitivity of stock prices to financial news (which are exogenous to markets) we might need to reconsider this rule. Recall that for some trading days we will have 400 observations for one asset, which will be very volatile. For example for Abbott Laboratories
(ABT), we have 20000 observations from 1 Dec. 2017 to 15 Feb. 2018. (Figure 1). The proper time interval to check performance of several trading experts might be last two hours time period for all Experts. So for this time period, I will prepare profitability of each Expert based on proposed trading signal of each expert and then calculate the return of each signal within this time period. 
This will be out of bag sample for all Experts. For Cross-validation of each expert, we need to consider the same size time period. More details of how to cross-validate the results of each Expert will be discussed later.  
\clearpage

\subsection{Times Series approach}
\subsubsection{ARIMA}
 ARIMA process simply, takes to account some lag of historical prices and weights them to predict today's prices. Although it might look simple, the question of number of lags to take into account is not an easy task. But it is common among scholars to select the proper lag by considering information criterion such as Akaike information criterion (AIC), and Hannan Quinn Information Criterion (HQIC). In this we assume stock log-returns (say $X_t$) are related to past prices. Then we can model:
 \begin{equation}
 X_t=\alpha_1 X_{t-1}+\alpha_2 X_{t-2}+...+\alpha_p X_{t-p}+\epsilon_t+\theta_1\epsilon_{t-1}+...\theta_q\epsilon_{t-q}
 \end{equation}
(p,q) are called lag parameters. Given that log-return prices follow stationary process (refer to appendix A for more detail.), we can estimate above procedure through MLE estimator. The proper lag parameters of ARIMA model is selected by AIC condition:
\begin{equation}
AIC= - Log(L)+2(p+q+k+1)
\end{equation}

For example, the ARIMA process applied to Abbott Laboratories log-returns , from: "2017-12-01" to: "2018-02-16 17:01:00", and the following result is obtained. This implies that we can model the mentioned stock price as follow
\begin{equation}
 X_t=0.000003274-0.0279X_{t-1}-0.00321 X_{t-1}-0.0384 X_{t-2}-0.0279X_{t-3}-0.0017X_{t-4}
 \end{equation}
 
Now according to above formula we can generate our forecast for future log-returns of underlying asset. For ARIMA process, I will keep 120 last observations for out of bag test and 120 as cross-validation of trading strategy.

The forecast results for Abbott Laboratories for 120 Minutes are depicted in following figure. Red line corresponds to  real price, and blue line depicts ARIMA(4,0,0) process forecast. If trading signals (as discussed in its section) are generated based on our forecast and allowing for short-selling in the market and getting advantage of selling strategy profits, ARIMA(4,0,0) will generate -\%16.62 loss for this asset within the 115 minutes trading time.  

 \begin{figure}[!htb]
 \frame{\includegraphics[width=14cm]{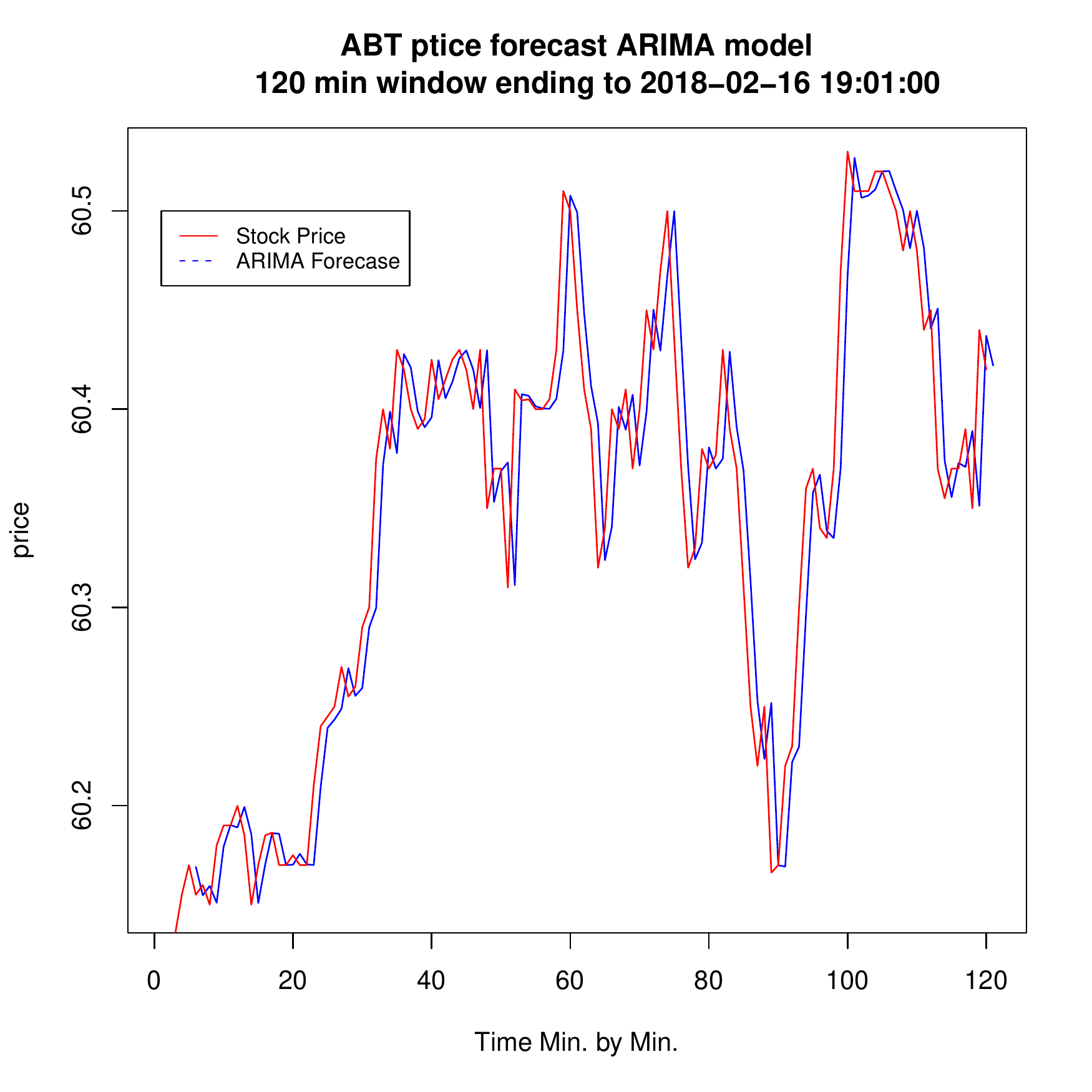}}
\label{Fig:Data13}  
\caption{Abbott Laboratories price, Online-learning ARIMA forecast}
\end{figure}
\begin{table}[ht]
\centering
\begin{tabular}{rrrlrr}
  \hline
 & today & forecast & position & tmw & porfit\_loss \\ 
  \hline
  4 & 60.15000 & 60.15101 & Long & 60.18000 & 0.03000 \\ 
  5 & 60.18000 & 60.17922 &  & 60.19000 & -0.00000 \\ 
  6 & 60.19000 & 60.19020 &  & 60.19000 & 0.00000 \\ 
  7 & 60.19000 & 60.18903 &  & 60.19990 & -0.00000 \\ 
  8 & 60.19990 & 60.19939 &  & 60.18500 & 0.00000 \\ 
  9 & 60.18500 & 60.18556 &  & 60.15000 & -0.00000 \\ 
  10 & 60.15000 & 60.15084 &  & 60.17000 & 0.00000 \\ 
  11 & 60.17000 & 60.17031 &  & 60.18500 & 0.00000 \\ 
  12 & 60.18500 & 60.18608 & Long & 60.18640 & 0.00140 \\ 
  13 & 60.18640 & 60.18580 &  & 60.17000 & 0.00000 \\ 
  14 & 60.17000 & 60.17004 &  & 60.17000 & 0.00000 \\ 
  15 & 60.17000 & 60.17017 &  & 60.17500 & 0.00000 \\ 
  16 & 60.17500 & 60.17568 &  & 60.17000 & -0.00000 \\ 
  17 & 60.17000 & 60.17035 &  & 60.17000 & 0.00000 \\ 
  18 & 60.17000 & 60.17002 &  & 60.21000 & 0.00000 \\ 
  19 & 60.21000 & 60.20926 &  & 60.24000 & -0.00000 \\ 
  20 & 60.24000 & 60.23924 &  & 60.24500 & -0.00000 \\ 
  21 & 60.24500 & 60.24342 & Short & 60.25000 & -0.00500 \\ 
  22 & 60.25000 & 60.24882 & Short & 60.27000 & -0.02000 \\ 
  23 & 60.27000 & 60.26938 &  & 60.25500 & 0.00000 \\ 
  24 & 60.25500 & 60.25535 &  & 60.26000 & 0.00000 \\ 
  25 & 60.26000 & 60.25933 &  & 60.29000 & -0.00000 \\ 
  26 & 60.29000 & 60.28989 &  & 60.30000 & -0.00000 \\ 
  27 & 60.30000 & 60.29965 &  & 60.37500 & -0.00000 \\ 
  28 & 60.37500 & 60.37191 & Short & 60.40000 & -0.02500 \\ 
  29 & 60.40000 & 60.39882 & Short & 60.38000 & 0.02000 \\ 
  30 & 60.38000 & 60.37778 & Short & 60.43000 & -0.05000 \\ 
  31 & 60.43000 & 60.42778 & Short & 60.42000 & 0.01000 \\ 
  32 & 60.42000 & 60.42104 & Long & 60.40000 & -0.02000 \\ 
  33 & 60.40000 & 60.39890 & Short & 60.39000 & 0.01000 \\ 
  34 & 60.39000 & 60.39084 &  & 60.39500 & 0.00000 \\ 
  35 & 60.39500 & 60.39587 &  & 60.42500 & 0.00000 \\ 
  36 & 60.42500 & 60.42476 &  & 60.40500 & 0.00000 \\ 
  37 & 60.40500 & 60.40548 &  & 60.41500 & 0.00000 \\ 
  38 & 60.41500 & 60.41382 & Short & 60.42500 & -0.01000 \\ 
  39 & 60.42500 & 60.42560 &  & 60.43000 & 0.00000 \\ 
  40 & 60.43000 & 60.42968 &  & 60.42000 & 0.00000 \\ 
  41 & 60.42000 & 60.42006 &  & 60.40000 & -0.00000 \\ 
  42 & 60.40000 & 60.40058 &  & 60.43000 & 0.00000 \\ 
  43 & 60.43000 & 60.42980 &  & 60.35000 & 0.00000 \\ 
  44 & 60.35000 & 60.35312 & Long & 60.37000 & 0.02000 \\ 
  45 & 60.37000 & 60.36878 & Short & 60.37000 & -0.00000 \\ 
   \hline
\end{tabular}
\end{table}
\hfill 
\clearpage
 \subsubsection{Kalman Filter}
 In this section linear State Space Model is considered to estimate future trading signals. To do so, I used Kalman filtering ,see \cite{kalman1960new} , method to estimate tomorrow's state of trade and then employed Expected Maximization Algorithm to estimate initial parameters for KF of each state. This will help to improve performance of KF in estimating trading signals. The State-Space model has two conditions: Hidden or latent process $x_t$ , state process, assumed to be a Markov process. The future $\{x_s; s > t\}$, and the past  past $\{x_s; s < t\}$, are independent conditional on the present, $x_t$. The observations, $y_t$ are independent given the states $x_t$. This means that the dependence among the observations is generated by states. Consider linear Gaussian state space model or (DLM), order one, p-dimensional vector autoregression as the state equation
\begin{equation}
x_t=\Phi x_{t-1}+w_t, \quad w_t \overset{iid}{\sim} N_p(0, Q) 
\end{equation}
Note that $w_t$ is a $p \times 1$ iid, zero-mean normal vectors with covariance matrix Q; the process starts with a normal vector $x_0$, such that $x_0 \sim N(\mu_0, \Sigma_0)$. We do not observe the state vector $x_t$ directly, but only a linear transformed
version of it with noise added, hence observation model, follows :
\begin{equation}
y_t=A_t x_t+v_t, \quad v_t \overset{iid}{\sim} N_q(0, R)
\end{equation}
$A_t$ is a $q \times p$ measurement or observation matrix.
The observed data vector, $y_t$, is q-dimensional, which can be larger than or
smaller than p, the state dimension.The additive observation noise is vector of Normally distributed data. For simplicity, $x_0, \{w_t\} and \{v_t\}$ are assumed to be uncorrelated. 
\begin{figure}[H] 
\centering
	 \frame{\includegraphics[width=0.6\textwidth] {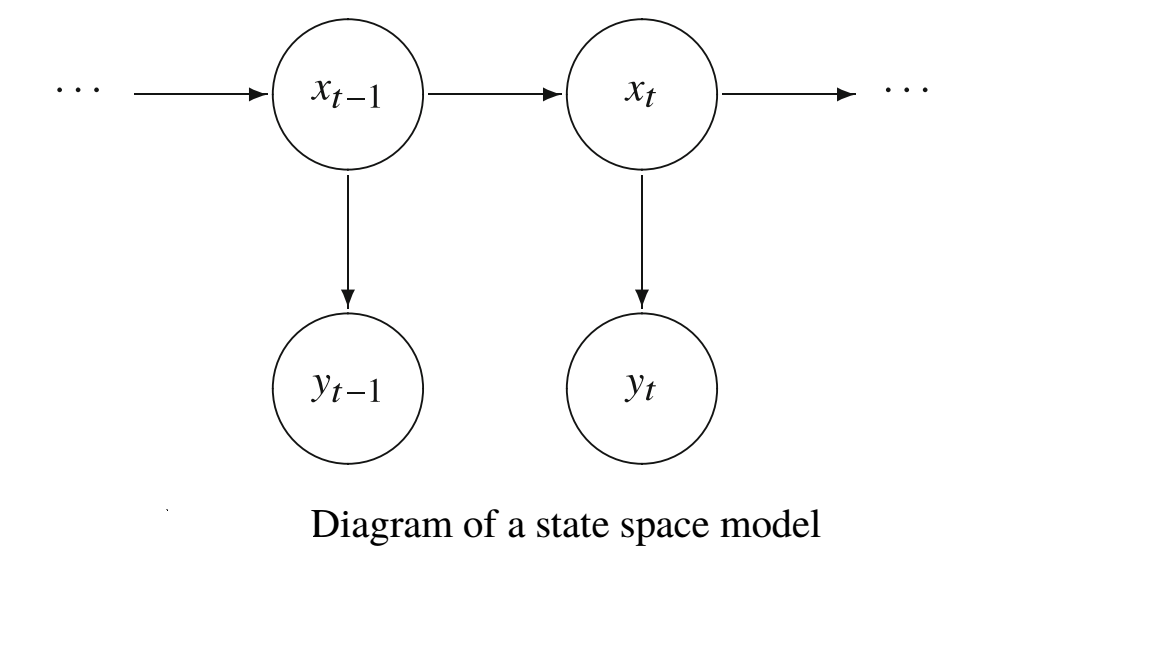}}
\label{fig:N1}  
\end{figure}

Our goal is to produce estimators for the underlying unobserved signal $x_t$, given the data $y_{1:s} = \{y_1, . . . , y_s\}$, to time s. When $s < t$, the problem is called forecasting or prediction. 
 When $s = t$, the problem is called filtering.
Given a sequence of observations $y_{1:t}$ the goal in stochastic filtering is to
compute the posterior distribution of the
internal states $p(x_t | y_{1:t})$. If X, Y process is jointly Gaussian, then the distribution of any subset of X conditioned on any subset of Y will
also be Gaussian. It follows that the filtering distributions are fully specified by
their means and variances. When $s > t$, the problem is called smoothing. Hence the goal is to compute $p(x_t | y_{1:s} )$ where $s > t$. This is a non-causal
process. So we use future observations to estimate the present state. Call: \\
$x_t^s = E(x_t | y_{1:s})$ \\

$P_{t_1,t_2}^s = E \{ (x_{t_1} - x_{t_1}^s)(x_{t_2} - x_{t_2}^s)'  | y_{1:s} \}$. \\
When $t_1 = t_2 (= t)$ then  $P_t^s$ is used for convenience.
$P_0^0 = \Sigma_0$ is initially known. Formula's to estimate Kalman Filter (\cite{kalman1960new} for the prove and more details):

\begin{eqnarray}
x_t^{t-1}=\Phi x_{t-1}^{t-1} \\
P_t^{t-1}=\Phi P_{t-1}^{t-1} \Phi'+Q \\
\Sigma_t=A_t P_t^{t-1}A'_t+R\\
K_t=P_t^{t-1} . A'_t[A_t P_t^{t-1}A'_t+R]^{-1}\\
x_t^t=x_t^{t-1}+K_t(y_t-A_t x_t^{t-1})\\
P_t^t=[I-K_t A_t]P_t^{t-1}\\
\epsilon_t = y_t - E(y_t | y_{1:t-1}) = y_t - A_t x_t^{t-1}
\end{eqnarray}

\subsubsection{Expectation Maximization algorithm}
\cite{shumway1982approach} proposed an EM algorithm conjunction with the Kalman Filter algorithm and derive a recursive procedure for estimating the parameters by MLE. The basic idea, is that if we know the states and observations, so we have complete data. So the joint distribution will be :

\begin{equation}
P_{\Theta} (x,y)=P(x0) \prod_{t=1}^n P(x_t|x_{t-1}) \prod_{t} P(y_t|x_{t})
\end{equation}

so the Log-likelihood will be :

\begin{eqnarray}
-2ln L_{X,Y}(\Theta)= ln (det(\Sigma_0))+(x_0 - \mu_0)'\Sigma_0^{-1} (x_0 - \mu_0)\\
+n ln (det (Q))+ \sum_{t=1}^n (x_t - \phi x_{t-1})'Q^{-1} (x_t - \phi x_{t-1})\\
+n ln (det (R))+ \sum_{t=1}^n (y_t - A_t x_{t})'R^{-1} (y_t - A_t x_{t})
\end{eqnarray}

Now, we need to construct the iteration steps. First we need expectation step:

\begin{eqnarray*}
E_j= ln (det(\Sigma_0))+tr (\Sigma_0^{-1} [P_0^n+ (x_0^n - \mu_0) (x_0^n - \mu_0)' )\\
+n ln (det (Q))+ tr( Q^{-1} [F_{11} -F{10} \phi'- \phi F'{10}+\phi F_{00} \phi' ) \\
+n ln (det (R))+ tr(R^{-1} \sum_{t=1}^n [(y_t - A_t x_{t}^n)' (y_t - A_t x_{t}^n)+A_tP_t^nA'_t] )
\end{eqnarray*}

where 
\begin{eqnarray}
F_{11}=\sum_{t=1}^n (x_t^n x_{t}^{n'}+P_t^n)\\
F_{10}=\sum_{t=1}^n (x_t^n x_{t-1}^{n'}+P_{t,t-1}^n)\\
F_{00}=\sum_{t=1}^n (x_{t-1}^n x_{t-1}^{n'}+P_{t-1}^n)
\end{eqnarray}

For min minimization step, one can derive the following updated estimations:

\begin{eqnarray}
\Phi_j=F_{10} F_{00}^{-1} \\
E_j=\frac{ F_{11}-F_{10} F_{00}^{-1} F'_{10} }{n}\\
R_j=\frac{\sum_{t-=1}^n (y_t - A_t x_{t}^n)' (y_t - A_t x_{t}^n)'+A_t P_t^n A'_t}{n} \\
\mu_0^j=x_0^n\\
\Sigma_o^j=P_0^n
\end{eqnarray}

Now, one can estimate above parameters by employing EM algorithm together with KF in each iteration. The corresponding code follows can be found in appendix. Now, by iterating above procedure, we get the estimations for our initial parameters.

Using above formula's, I estimated state-space model and plotted for ABT firm in following graph. (for detail of codes please refer to Appendix). The green plot demonstrates the minute by minute stock prices and the red ones is prediction by the model. Kalman Filter can be considered as an online learning method. Employing this method on 265 minutes (test data), profits 0.89 per share. Trading strategy was taking long position in case of our prediction for tomorrow price is higher than today's price and taking short position if we predict price will fall tomorrow. The next day, the current position will be closed, no matter what happens. (Here tomorrow means next minute as I have used minutes by minutes data).   

 \begin{figure}[!htb]
 \frame{\includegraphics[width=16cm]{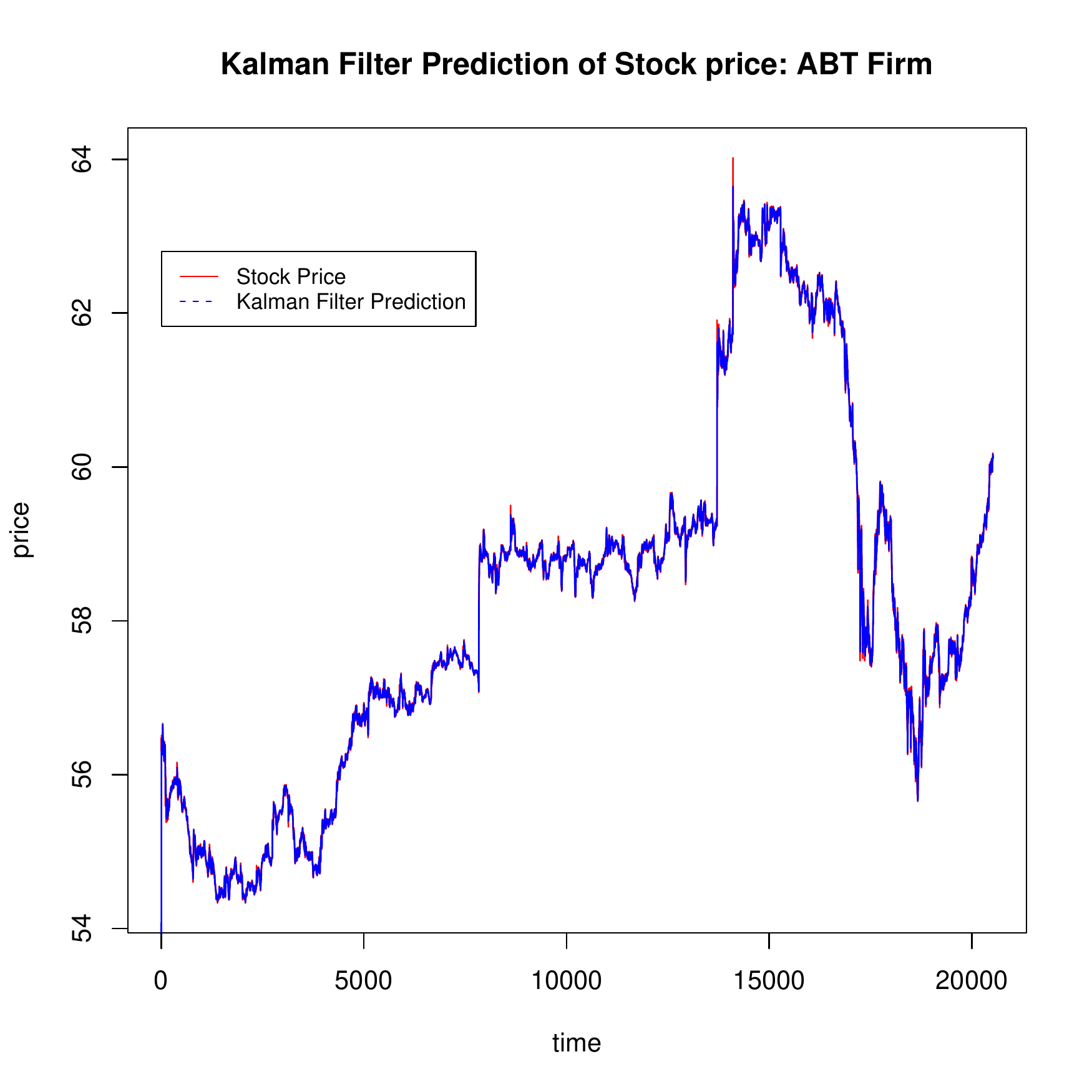}}
\label{Fig:Data7}  
\caption{Abbott Laboratories Stock prices prediction with Kalman filter }
\end{figure}
\hfill 

 \begin{figure}[!htb]
 \frame{\includegraphics[width=16cm]{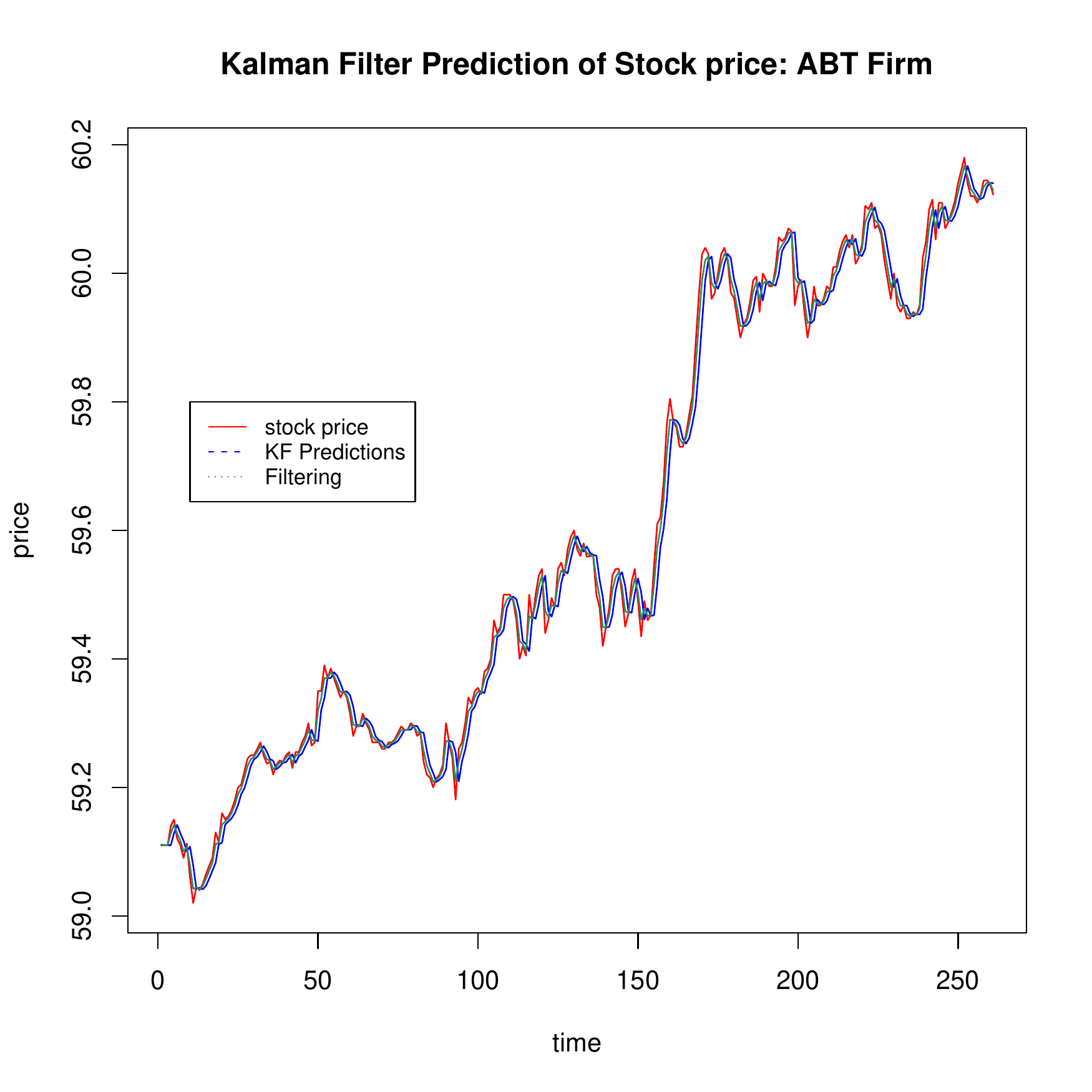}}
\label{Fig:Data8}  
\caption{Abbott Laboratories Stock prices prediction with Kalman filter- for 264 minutes }
\end{figure}
\hfill

\begin{table}[ht]
\centering
\begin{tabular}{rrrlrr}
  \hline
 & today & forecast & position & tmw & porfit\_loss \\ 
  \hline
1 & 59.11000 & 59.11086 &  & 59.11000 & 0.00000 \\ 
  2 & 59.11000 & 59.11033 &  & 59.11000 & 0.00000 \\ 
  3 & 59.11000 & 59.11012 &  & 59.14000 & 0.00000 \\ 
  4 & 59.14000 & 59.11005 & Short & 59.15000 & -0.01000 \\ 
  5 & 59.15000 & 59.12856 & Short & 59.12000 & 0.03000 \\ 
  6 & 59.12000 & 59.14181 & Long & 59.11000 & -0.01000 \\ 
  7 & 59.11000 & 59.12833 & Long & 59.09000 & -0.02000 \\ 
  8 & 59.09000 & 59.11700 & Long & 59.11260 & 0.02260 \\ 
  9 & 59.11260 & 59.10031 & Short & 59.06000 & 0.05260 \\ 
  10 & 59.06000 & 59.10791 & Long & 59.02000 & -0.04000 \\ 
  11 & 59.02000 & 59.07830 & Long & 59.04500 & 0.02500 \\ 
  12 & 59.04500 & 59.04227 & Short & 59.04000 & 0.00500 \\ 
  13 & 59.04000 & 59.04396 & Long & 59.05000 & 0.01000 \\ 
  14 & 59.05000 & 59.04151 & Short & 59.06500 & -0.01500 \\ 
  15 & 59.06500 & 59.04676 & Short & 59.07800 & -0.01300 \\ 
  16 & 59.07800 & 59.05803 & Short & 59.09000 & -0.01200 \\ 
  17 & 59.09000 & 59.07037 & Short & 59.13000 & -0.04000 \\ 
  18 & 59.13000 & 59.08250 & Short & 59.11500 & 0.01500 \\ 
  19 & 59.11500 & 59.11186 & Short & 59.16000 & -0.04500 \\ 
  20 & 59.16000 & 59.11380 & Short & 59.15000 & 0.01000 \\ 
  21 & 59.15000 & 59.14235 & Short & 59.15500 & -0.00500 \\ 
  22 & 59.15500 & 59.14708 & Short & 59.16500 & -0.01000 \\ 
  23 & 59.16500 & 59.15197 & Short & 59.18000 & -0.01500 \\ 
  24 & 59.18000 & 59.16002 & Short & 59.20000 & -0.02000 \\ 
  25 & 59.20000 & 59.17237 & Short & 59.20440 & -0.00440 \\ 
  26 & 59.20440 & 59.18945 & Short & 59.22500 & -0.02060 \\ 
  27 & 59.22500 & 59.19869 & Short & 59.24500 & -0.02000 \\ 
  28 & 59.24500 & 59.21495 & Short & 59.25000 & -0.00500 \\ 
  29 & 59.25000 & 59.23352 & Short & 59.25000 & -0.00000 \\ 
  30 & 59.25000 & 59.24371 & Short & 59.26000 & -0.01000 \\ 
  31 & 59.26000 & 59.24760 & Short & 59.27000 & -0.01000 \\ 
  32 & 59.27000 & 59.25526 & Short & 59.25000 & 0.02000 \\ 
  33 & 59.25000 & 59.26437 & Long & 59.23680 & -0.01320 \\ 
  34 & 59.23680 & 59.25549 & Long & 59.24000 & 0.00320 \\ 
  35 & 59.24000 & 59.24394 & Long & 59.22000 & -0.02000 \\ 
  36 & 59.22000 & 59.24150 & Long & 59.23500 & 0.01500 \\ 
  37 & 59.23500 & 59.22821 & Short & 59.24200 & -0.00700 \\ 
  38 & 59.24200 & 59.23241 & Short & 59.24000 & 0.00200 \\ 
  39 & 59.24000 & 59.23834 & Short & 59.25000 & -0.01000 \\ 
  40 & 59.25000 & 59.23936 & Short & 59.25500 & -0.00500 \\ 
  41 & 59.25500 & 59.24594 & Short & 59.23000 & 0.02500 \\ 
  42 & 59.23000 & 59.25154 & Long & 59.25500 & 0.02500 \\ 
  43 & 59.25500 & 59.23823 & Short & 59.25500 & -0.00000 \\ 
  44 & 59.25500 & 59.24859 & Short & 59.27000 & -0.01500 \\ 
  45 & 59.27000 & 59.25255 & Short & 59.28000 & -0.01000 \\ 
  46 & 59.28000 & 59.26334 & Short & 59.30000 & -0.02000 \\ 
  47 & 59.30000 & 59.27363 & Short & 59.26500 & 0.03500 \\ 
  48 & 59.26500 & 59.28993 & Long & 59.27000 & 0.00500 \\ 
 
   \hline
\end{tabular}
\end{table}

\clearpage
\subsection{Machine Learning Algorithms}

\subsubsection{Neural Network}
\cite{patel2015predicting} and \cite{booth2015performance} compared different ML algorithms which are applied to predict stock price movements. They compared ensemble of random forests, with Neural Network, Support Vector Regression, and OLS, and found the superior results of the former over other methods. In this paper, I will compare performance of these methods: Ensemble of Random Forest, in contrast to Times series data. Although \cite{langkvist2014review} survey, is hopeless on employing NN methods on stock market prediction, due to its non-stationarity movements, but there might be a better chance of trying NN with predicting trading signals (which, as discussed,is bounded time series) as an extension of this research.

\subsubsection{Random Forest}
 
 \cite{creamer2010automated} and \cite{patel2015predicting} report that among several Machine learning algorithms, ensemble Random Forests perform better to predict direction of stock prices. Random forests (RF) algorithm introduced by \cite{breiman2001random} as an average of some random regression trees to avoid over-fitting of each tree. Although some scholars implemented RF successfully to predict financial markets (\cite{kara2011predicting}), but problem arises when data is time dependent, or sequential data. The novelty of this research on predicting trading signal might have some positive impact to solve this problem. 
  \begin{figure}[!htb]
  \centering
 \frame{\includegraphics[width=10cm]{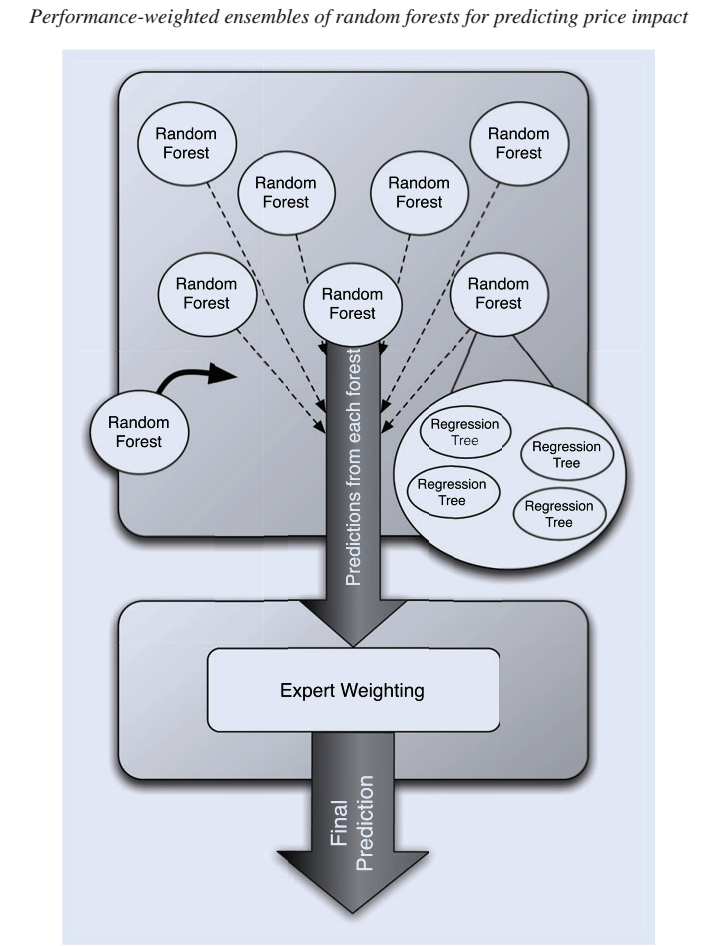}}
\label{Fig:Data10}  
\caption{Ensemble of Random Forests }
\end{figure}
\hfill 
 Also, I will work on new weighting algorithm to combine several RF. This is normally called Ensemble of RF in literature. \cite{saffari2009line} proposed online Random Forest algorithm which is based on the online bagging, randomized forests and online regression tree. Moreover, they implemented a weighting algorithm to punish some regression trees based on their out-of-bag-error to grow new regression trees. This idea looks similar to Adaboost learning algorithm of \cite{freund1995desicion}, which repeatedly uses a
simple learning algorithm, called weak learner to different weightings of training data. While Adaboost performs relatively good for predicting direction of  financial markets, one might be interested to see its performance on magnitude of future prices. Some scholars combined the Adaboost algorithm with RF algorithm in this manner. First they trained several RF on time intervals of time series data, then considered each RF an weak expert to predict tomorrow prices. Then they punished weak RF experts based on the outcomes of each expert to predict prices (see \cite{creamer2010automated}. In this study, I used this idea, but with some modifications on weighting algorithm of each weak RF. Each RF will be applied to some interval of data. Then, I will weight each RF based on inverse of their absolute prediction error. Then will update the weights of each RF, to predict tomorrow's (next minutes here) trading signals based on ensemble of weighted Random Forests. After that, I will try to benefit from new Minutes by Minutes data and update each prediction by correcting previous predictions measures. So for each next Min. we will best use of today's information. In the following plot, I illustrated the the results of RF on test data set of Minutes by Minutes prices for ABT company. Next, I have considered our predictions as trading signals and check the profitability of proposed algorithm. For now, the algorithm has 7.5 \% success in direction estimation. 

 \begin{figure}[!htb]
 \frame{\includegraphics[width=16cm]{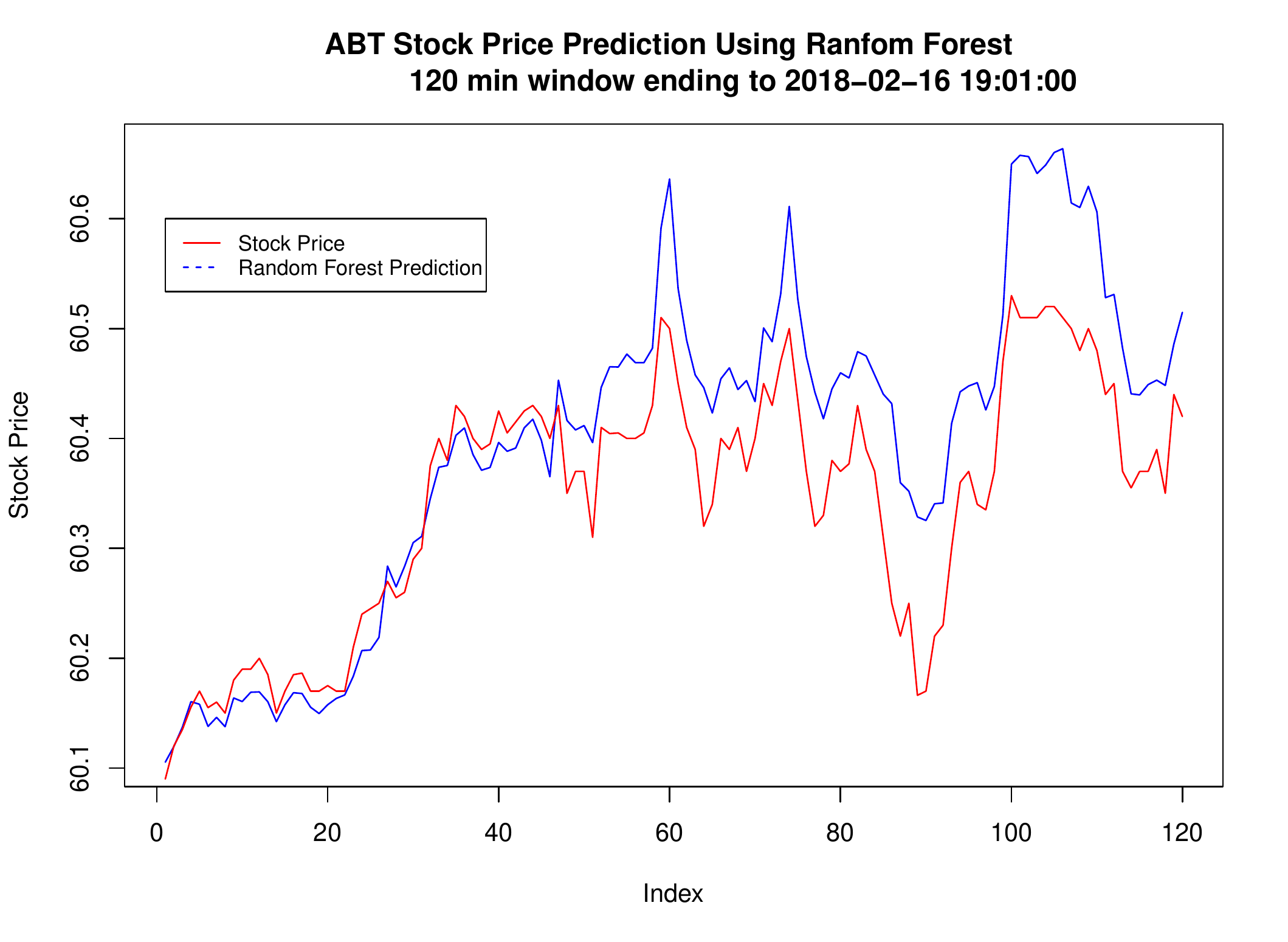}}
\label{Fig:Data4}  
\caption{Abbott Laboratories Stock prices prediction with Random Forest }
\end{figure}
\hfill 
\clearpage
\begin{table}[ht]
\centering
\begin{tabular}{rrrrlrr}
  \hline
 & row & today.price & forecast.tmw & signal & tmw & profit.loss \\ 
  \hline
1 &     1 & 60.0900 & 60.1054 & Long & 60.1200 & 0.0300 \\ 
  2 &     2 & 60.1200 & 60.1196 &  & 60.1350 & 0.0000 \\ 
  3 &     3 & 60.1350 & 60.1371 &  & 60.1554 & 0.0000 \\ 
  4 &     4 & 60.1554 & 60.1604 &  & 60.1700 & 0.0000 \\ 
  5 &     5 & 60.1700 & 60.1581 & Short & 60.1550 & 0.0150 \\ 
  6 &     6 & 60.1550 & 60.1379 & Short & 60.1600 & -0.0050 \\ 
  7 &     7 & 60.1600 & 60.1461 & Short & 60.1500 & 0.0100 \\ 
  8 &     8 & 60.1500 & 60.1376 & Short & 60.1800 & -0.0300 \\ 
  9 &     9 & 60.1800 & 60.1638 & Short & 60.1900 & -0.0100 \\ 
  10 &    10 & 60.1900 & 60.1605 & Short & 60.1900 & 0.0000 \\ 
  11 &    11 & 60.1900 & 60.1690 & Short & 60.1999 & -0.0099 \\ 
  12 &    12 & 60.1999 & 60.1693 & Short & 60.1850 & 0.0149 \\ 
  13 &    13 & 60.1850 & 60.1603 & Short & 60.1500 & 0.0350 \\ 
  14 &    14 & 60.1500 & 60.1422 &  & 60.1700 & 0.0000 \\ 
  15 &    15 & 60.1700 & 60.1576 & Short & 60.1850 & -0.0150 \\ 
  16 &    16 & 60.1850 & 60.1686 & Short & 60.1864 & -0.0014 \\ 
  17 &    17 & 60.1864 & 60.1678 & Short & 60.1700 & 0.0164 \\ 
  18 &    18 & 60.1700 & 60.1553 &  & 60.1700 & 0.0000 \\ 
  19 &    19 & 60.1700 & 60.1496 & Short & 60.1750 & -0.0050 \\ 
  20 &    20 & 60.1750 & 60.1576 & Short & 60.1700 & 0.0050 \\ 
  21 &    21 & 60.1700 & 60.1633 &  & 60.1700 & 0.0000 \\ 
  22 &    22 & 60.1700 & 60.1666 &  & 60.2100 & 0.0000 \\ 
  23 &    23 & 60.2100 & 60.1836 & Short & 60.2400 & -0.0300 \\ 
  24 &    24 & 60.2400 & 60.2069 & Short & 60.2450 & -0.0050 \\ 
  25 &    25 & 60.2450 & 60.2074 & Short & 60.2500 & -0.0050 \\ 
  26 &    26 & 60.2500 & 60.2189 & Short & 60.2700 & -0.0200 \\ 
  27 &    27 & 60.2700 & 60.2838 & Long & 60.2550 & -0.0150 \\ 
  28 &    28 & 60.2550 & 60.2648 &  & 60.2600 & 0.0000 \\ 
  29 &    29 & 60.2600 & 60.2836 & Long & 60.2900 & 0.0300 \\ 
  30 &    30 & 60.2900 & 60.3051 & Long & 60.3000 & 0.0100 \\ 
  31 &    31 & 60.3000 & 60.3109 & Long & 60.3750 & 0.0750 \\ 
 
   \hline
\end{tabular}
\end{table}
\clearpage
\subsubsection{Other Prediction Models}
\cite{hochreiter1997long} proposed Long Short-term Memory (LSTM) Networks, which are a type of recurrent neural networks and applied successfully to predict stock markets data. It will be worth to see the performance of LSTM method to predict the trading signals of stock prices, in future research.

\section{Conclusion}
The goal of this report is to compare the time series methods versus some machine learning based algorithms for stock price movements and generating trading signals. The use of on-line learning algorithms is forced by nature of stock markets price behavior, as they are prone to exogenous shocks. To avoid effect of these shocks (some news shocks and some macro-foundations shocks) using on-line learning algorithms will be helpful. The study shows, random forest algorithm performs considerably better than Arima (which is simply regressing today's prices on history of prices). But employing Kalman filter together with Expectation Maximization algorithm, results way better than Ensemble of Random Forest Algorithm. One might need to check if Kalman Filter performs better than other Machine Learning algorithms such as Long-Short Term Memory in future research.

\newpage

\bibliographystyle{unsrt}  
\bibliography{references.bib}  

\begin{thebibliography}{10}

\bibitem{murphy1999technical}
John~J Murphy.
\newblock {\em Technical analysis of the financial markets: A comprehensive
  guide to trading methods and applications}.
\newblock Penguin, 1999.

\bibitem{teixeira2010method}
Lamartine~Almeida Teixeira and Adriano Lorena~Inacio De~Oliveira.
\newblock A method for automatic stock trading combining technical analysis and
  nearest neighbor classification.
\newblock {\em Expert systems with applications}, 37(10):6885--6890, 2010.

\bibitem{godfrey1964random}
Michael~D Godfrey, Clive~WJ Granger, and Oskar Morgenstern.
\newblock The random-walk hypothesis of stock market behavior.
\newblock {\em Kyklos}, 17(1):1--30, 1964.

\bibitem{fama1965behavior}
Eugene~F Fama.
\newblock The behavior of stock-market prices.
\newblock {\em The journal of Business}, 38(1):34--105, 1965.

\bibitem{bondt1985does}
Werner~FM Bondt and Richard Thaler.
\newblock Does the stock market overreact?
\newblock {\em The Journal of finance}, 40(3):793--805, 1985.

\bibitem{jensen1978some}
Michael~C Jensen.
\newblock Some anomalous evidence regarding market efficiency.
\newblock {\em Journal of financial economics}, 6(2-3):95--101, 1978.

\bibitem{lui1998use}
Yu-Hon Lui and David Mole.
\newblock The use of fundamental and technical analyses by foreign exchange
  dealers: Hong kong evidence.
\newblock {\em Journal of International Money and Finance}, 17(3):535--545,
  1998.

\bibitem{creamer2010automated}
Germ{\'a}n Creamer and Yoav Freund.
\newblock Automated trading with boosting and expert weighting.
\newblock {\em Quantitative Finance}, 10(4):401--420, 2010.

\bibitem{kara2011predicting}
Yakup Kara, Melek~Acar Boyacioglu, and {\"O}mer~Kaan Baykan.
\newblock Predicting direction of stock price index movement using artificial
  neural networks and support vector machines: The sample of the istanbul stock
  exchange.
\newblock {\em Expert systems with Applications}, 38(5):5311--5319, 2011.

\bibitem{booth2015performance}
Ash Booth, Enrico Gerding, and Frank McGroarty.
\newblock Performance-weighted ensembles of random forests for predicting price
  impact.
\newblock {\em Quantitative Finance}, 15(11):1823--1835, 2015.

\bibitem{dickey1979distribution}
David~A Dickey and Wayne~A Fuller.
\newblock Distribution of the estimators for autoregressive time series with a
  unit root.
\newblock {\em Journal of the American statistical association},
  74(366a):427--431, 1979.

\bibitem{gately1995neural}
Edward Gately.
\newblock {\em Neural networks for financial forecasting}.
\newblock John Wiley \& Sons, Inc., 1995.

\bibitem{cavalcante2016computational}
Rodolfo~C Cavalcante, Rodrigo~C Brasileiro, Victor~LF Souza, Jarley~P Nobrega,
  and Adriano~LI Oliveira.
\newblock Computational intelligence and financial markets: A survey and future
  directions.
\newblock {\em Expert Systems with Applications}, 55:194--211, 2016.

\bibitem{kalman1960new}
Rudolph~Emil Kalman.
\newblock A new approach to linear filtering and prediction problems.
\newblock {\em Journal of basic Engineering}, 82(1):35--45, 1960.

\bibitem{shumway1982approach}
Robert~H Shumway and David~S Stoffer.
\newblock An approach to time series smoothing and forecasting using the em
  algorithm.
\newblock {\em Journal of time series analysis}, 3(4):253--264, 1982.

\bibitem{patel2015predicting}
Jigar Patel, Sahil Shah, Priyank Thakkar, and K~Kotecha.
\newblock Predicting stock and stock price index movement using trend
  deterministic data preparation and machine learning techniques.
\newblock {\em Expert Systems with Applications}, 42(1):259--268, 2015.

\bibitem{langkvist2014review}
Martin L{\"a}ngkvist, Lars Karlsson, and Amy Loutfi.
\newblock A review of unsupervised feature learning and deep learning for
  time-series modeling.
\newblock {\em Pattern Recognition Letters}, 42:11--24, 2014.

\bibitem{breiman2001random}
Leo Breiman.
\newblock Random forests.
\newblock {\em Machine learning}, 45(1):5--32, 2001.

\bibitem{saffari2009line}
Amir Saffari, Christian Leistner, Jakob Santner, Martin Godec, and Horst
  Bischof.
\newblock On-line random forests.
\newblock In {\em Computer Vision Workshops (ICCV Workshops), 2009 IEEE 12th
  International Conference on}, pages 1393--1400. IEEE, 2009.

\bibitem{freund1995desicion}
Yoav Freund and Robert~E Schapire.
\newblock A desicion-theoretic generalization of on-line learning and an
  application to boosting.
\newblock In {\em European conference on computational learning theory}, pages
  23--37. Springer, 1995.

\bibitem{hochreiter1997long}
Sepp Hochreiter and J{\"u}rgen Schmidhuber.
\newblock Long short-term memory.
\newblock {\em Neural computation}, 9(8):1735--1780, 1997.

\end{thebibliography}



\end{document}